\documentclass[aip,reprint]{revtex4-2}

\usepackage[T1]{fontenc}
\usepackage[utf8]{inputenc}
\usepackage{graphicx}
\usepackage[english]{babel}
\usepackage{upgreek}
\usepackage{color}
\usepackage{amsmath}
\usepackage{textcomp}
\usepackage[
colorlinks=true,
linkcolor=blue,
citecolor=blue,
urlcolor = blue]{hyperref}

\draft 

\begin{document}


\title{\color{blue}
Three-dimensional magnetic resonance tomography with sub-10 nanometer resolution 
}



\author{Mohammad T Amawi}

\affiliation{Institute for Physics, Quantum technology, University of Rostock, Germany}
\affiliation{Physics Department, Technical University of Munich, Germany}
\affiliation{Department of Life, Light and Matter, Quantum Technology, University of Rostock,Germany}
\affiliation{Munich Center for Quantum Science and Technology (MCQST), Germany}

\author{Andrii Trelin}
\affiliation{Institute for Physics, Quantum technology, University of Rostock, Germany}
\affiliation{Department of Life, Light and Matter, Quantum Technology, University of Rostock,Germany}

\author{You Huang}
\affiliation{Institute for Physics, Quantum technology, University of Rostock, Germany}
\affiliation{Department of Life, Light and Matter, Quantum Technology, University of Rostock,Germany}
\affiliation{CAS Key Laboratory of Microscale Magnetic Resonance and School of Physical Sciences, University of Science and Technology of China, Hefei 230026, China}
\affiliation{CAS Center for Excellence in Quantum Information and Quantum Physics, University of Science and Technology of China, Hefei 230026, China}

\author{Paul Weinbrenner}
\affiliation{Institute for Physics, Quantum technology, University of Rostock, Germany}
\affiliation{Department of Life, Light and Matter, Quantum Technology, University of Rostock,Germany}

\author{Francesco Poggiali}
\affiliation{Institute for Physics, Quantum technology, University of Rostock, Germany}
\affiliation{Department of Life, Light and Matter, Quantum Technology, University of Rostock,Germany}
\affiliation{Munich Center for Quantum Science and Technology (MCQST), Germany}

\author{Joachim Leibold}
\affiliation{Physics Department, Technical University of Munich, Germany}
\affiliation{Munich Center for Quantum Science and Technology (MCQST), Germany}

\author{Martin Schalk}
\affiliation{Physics Department, Technical University of Munich, Germany}
\affiliation{Munich Center for Quantum Science and Technology (MCQST), Germany}


\author{Friedemann Reinhard}
\email{friedemann.reinhard@uni-rostock.de}
\affiliation{Institute for Physics, Quantum technology, University of Rostock, Germany}
\affiliation{Department of Life, Light and Matter, Quantum Technology, University of Rostock,Germany}
\affiliation{Munich Center for Quantum Science and Technology (MCQST), Germany}


\date{\today}

\begin{abstract}
We demonstrate three-dimensional magnetic resonance tomography with a resolution down to $5.99\pm 0.07~$nm. Our measurements use lithographically fabricated microwires as a source of three-dimensional magnetic field gradients, which we use to image NV centers in a densely doped diamond by Fourier-accelerated magnetic resonance tomography. We also present a compressed sensing scheme for imaging of a spatially localized ensemble from undersampled data, which allows for a direct visual interpretation without numerical optimization. The resolution achieved in our work approaches the positioning accuracy of site-directed spin labeling, paving the way to three-dimensional structure analysis by magnetic-gradient based tomography. 
\end{abstract}

\maketitle 

In recent years, various nano-sensors, most prominently magnetic resonance force microscopy (MRFM) and nitrogen-vacancy (NV) centers, have enabled the detection of small ensembles of electron\cite{rugar04, shi15} and nuclear\cite{degen09, staudacher13, mamin13} spins, partially down to the level of single spins. Translating this power from a mere detection to a three-dimensional imaging technique promises transformative applications. Three-dimensional imaging of color centers would enable selective addressing and readout of networks of coherently coupled color centers in densely doped samples\cite{zhang17, artzi23}, or the detection of elementary particles by high-resolution mapping of the crystal strain they induce upon impact\cite{rajendran17}.
Applied to electron spins, it would enable three-dimensional imaging of spin-labeled proteins. Such a technique would in particular provide distance constraints for label distances of $>80~\textup{\AA}$ 
and for proteins labeled with arbitrary many electron spins, 
filling two blind spots of present electron spin resonance spectroscopy. Applied to nuclear spins, it would provide an ultimate microscope, able to image within opaque samples with label-free chemical contrast. 

Conceptually, the step from detection to imaging is straightforward. Applying a magnetic field gradient is all it takes to turn a magnetic resonance spectrum into a one-dimensional image. Multiple gradients along linearly independent directions can encode multi-dimensional images, most beautifully illustrated in the output of clinical magnetic resonance imaging scanners. If the gradients can be switched faster than the duration of the spectroscopy sequence, Fourier-accelerated techniques can acquire extended volumes in reasonable time\cite{arai15}. In clinical scanners, these comprise thousands of voxels, and similar data volumes are expected for particle detectors or imaging of a densely spin-labeled protein. Yet, three-dimensional Fourier-accelerated imaging at the nanoscale has remained elusive. 

Imaging by less scalable techniques has been demonstrated multiple times. Three-dimensional imaging of nuclear spins with atomic resolution has been achieved using the intrinsic field gradient emerging from the magnetic dipole field of a color center\cite{zopes18, abobeih19}. However, only the closest few nanometers around a defect can be imaged by this approach so that it is limited to intrinsic spins in the diamond so far. One-dimensional and two-dimensional\cite{dasilvabarbosa21} imaging in static gradients has been demonstrated, including resolving two adjacent centers by the gradient field of a hard-drive write head\cite{bodenstedt18}. Three-dimensional images of intrinsic electron spins in a diamond have been obtained using a static gradient positioned by a scanning probe\cite{grinolds14}. Fourier acceleration has remained out of reach of these approaches where gradients cannot be switched within a spectroscopy sequence. 

Fourier acceleration of imaging has been demonstrated\cite{arai15, artzi23} using quickly switchable conductors as gradient sources, but experiments with nanoscale resolution have remained limited to one- and two-dimensional proofs of the concept. One-dimensional imaging in MRFM by current-driven gradients has very recently even achieved sub-Ångstrom resolution\cite{haas22}. 




\begin{figure}
    \includegraphics[scale=1]{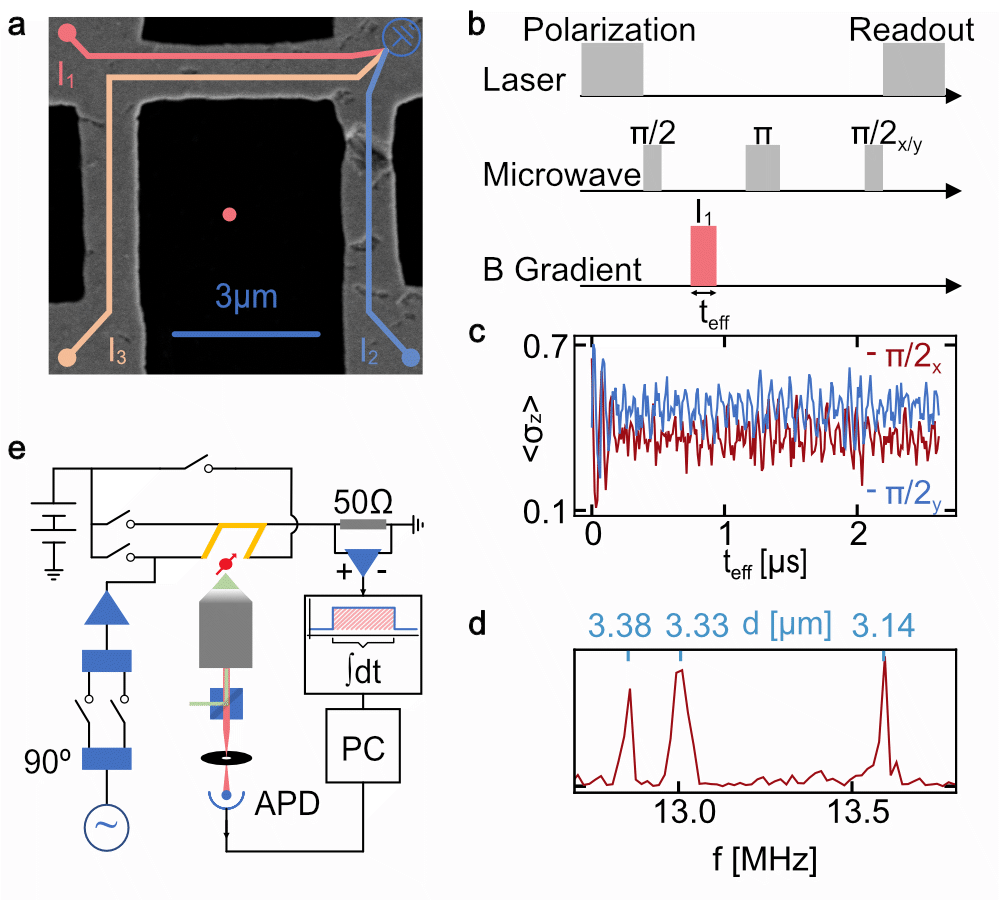}
    \caption{\label{fig:exp1} Experimental setup and one-dimensional magnetic resonance tomography of NV centers. (a) Electron micrograph of a device as used in the present study. Currents in the three gold wires of a microfabricated U-Structure create three linearly independent magnetic field gradients in the densely doped diamond below the structure. (b) Pulse sequence for one-dimensional imaging. A magnetic gradient pulse (length $t_{\textit{eff}}$) inserted into a Hahn echo sequence phase-encodes position. The NV spin state is initialized and the spin projection $\langle \hat S_z\rangle$ is read out optically. (c) Measurement result of (b). $\pi/2_x$ and $\pi/2_y$ denote the phase of the trailing $\pi/2$ pulse in (b). (d) Fourier transform of a dataset like (c) extending to $t=60~$\textmu s. Every NV center gives rise to one peak at the Larmor frequency set by the magnetic field $B_{I_1}(d)$ of the wire. $d$ denotes the distance from wire $I_1$. (e) Experimental setup. A single microwave generator, a $90^{\circ}$ splitter and two microwave switches are used to implement the Hahn Echo sequence. A confocal microscope with an avalanche photodiode (APD) as a detector is used for NV center polarization and readout. The gradient currents $I_1, I_2, I_3$ are created from a constant voltage source and can be pulsed by a fast switch. The voltage drop across the resistor is recorded by an A/D-converter and the pulse integral $\int I dt$ is saved for every single current pulse.}
\end{figure}

Here we demonstrate Fourier-accelerated three-dimensional imaging with nanometer-scale resolution. The key is a device to produce three linearly independent magnetic field gradients from a two-dimensional layout of conductors (Fig.~\ref{fig:exp1}a). Three microfabricated wires, arranged in a U-shape structure, create linearly independent gradient fields in a plane few microns beneath the structure. 
This device is fabricated via lift-off photolithography on a diamond substrate hosting a dense ($[\textit{NV}]\approx 0.13~$ppb) ensemble of NV centers Fig.~\ref{fig:exp1}(a). The U-microstructure consists of a $200~$nm gold film on top of a $10~$nm thick titanium layer. Each of its three arms is 5$~$\textmu m long and $500~$nm wide. The top arm also serves as a microwave antenna to implement single-qubit gates. We generate switchable magnetic field gradients by sending currents, labeled I$_1$, I$_2$, and I$_3$ in Fig.~\ref{fig:exp1}(a), into the three arms of the U microstructure. All currents are terminated with a $50~\Omega$ resistance at the same vertex of the U structure. In all measurements a homogeneous bias magnetic field of $B_0 \approx 76~$G is applied along one of the four NV axes.
This device is used to implement gradient echo pulse sequences, like the one-dimensional example shown in Fig.~\ref{fig:exp1}(b-d). A Hahn echo sequences decouples the NV centers from static and slowly fluctuating background fields, to enable $T_2$-limited sensing. A magnetic field gradient pulse, created by the current $I_1$, is applied during one half of the echo sequence. This phase-encodes the position, because an NV center at point $\vec x$ acquires a position-dependent phase shift
\begin{equation}
  \phi(\vec x, t) = \int_0^t \omega(\vec B_I(\vec x, \tau)) d\tau \approx \omega(\vec B_I(\vec x))t  
\label{eq:phase_encoding}
\end{equation}
where $\omega(\vec B_I(\vec x, \tau))$ denotes the shift in the Larmor frequency induced by the current $I$ and the approximation holds for pulses close to a rectangular shape. 
At the end of the Hahn echo sequence, this phase shift translates into an oscillatory spin signal 

 \[ \langle \hat S_z\rangle (t) = \begin{cases} (1 + \cos(\omega(\vec x) t)/2 & \text{ trailing $\pi_x/2$ pulse}  \\ (1 + \sin(\omega(\vec x) t)/2 & \text{ trailing $\pi_y/2$ pulse} \\ \end{cases}\]

For a distribution of NV centers, the oscillatory signals of all centers will linearly superpose to a characteristic beating pattern
$$
\langle \hat S_z\rangle (t) = \Bigg{(}1 + \sum_{\vec x_{\textit{NV}}}  \cos(\omega(\vec x_{\textit{NV}}) t)\Bigg{)}/2 
$$
(assuming a trailing $\pi_x/2$ pulse). 
The various $\omega(\vec x_{\textit{NV}})$ can be recovered from this signal by an inverse Fourier transform, creating a one-dimensional image of the NV centers Fig.~\ref{fig:exp1}(d). \par 
One major challenge of this experiment consists in creating sufficiently rectangular pulses to satisfy the approximation of eq. (\ref{eq:phase_encoding}). This requires a stable current supply that moreover has to be controlled with a fast ($100~$MHz) bandwidth to ensure that the rising and falling edges are quasi-instantaneous, i.e. much shorter than one period of the current-induced Larmor frequency $\omega(\vec B_I)$. Stability within every current pulse is required, because any variation of $\omega(\vec B(\vec x, \tau))$ over the pulse will introduce a chirp in the time domain signal (Fig.~\ref{fig:exp1}(c)), which will blur the image in the frequency domain (Fig.~\ref{fig:exp1}(d)). Stability between successive experimental repetitions is required, because shot-to-shot fluctuations of the magnetic field induce decoherence (see below). \par 
We experimentally address these constraints by two means (Fig.~\ref{fig:exp1}(e)). First, the current pulses are generated by switching a stable voltage source (Keithley 2230G-30-6) using fast switches (ic-Haus HGP), ensuring nearly rectangular pulses. Second, we correct for residual nonlinearities and fluctuations by measurement and online post-processing. We acquire the current integral $\int_0^t I(\tau) d\tau$ for every pulse of every experimental repetition by hardware-integration on a fast A/D converter (Spectrum M4i.4451-x8), and use this value to define a new time axis for all photonic measurements that removes chirps. Specifically, we make the following approximation (valid for nearly rectangular pulses and a nearly linear Zeeman shift)
$$\int_0^t \omega(\vec B_I(\vec x, \tau)) d\tau \approx \omega_{I, \textit{ref}}\frac{\int_0^t I(\tau) d\tau}{I_0} = \omega_{I, \textit{ref}}t_{\textit{eff}}$$
Here, $\omega_{I, \textit{ref}}(\vec x)$ is the shift in Larmor frequency induced by some reference current $I_0$ and $t_{\textit{eff}}$ denotes an "effective pulse duration". We thus absorb minor fluctuations of the current over the pulse into this redefined time coordinate $t_{\textit{eff}}$. All time-domain plots in this paper will use $t_{\textit{eff}}$ as time axis unless noted otherwise. 
This correction also suppresses shot-to-shot fluctuations, improving coherence\cite{braunbeck21}.

\begin{figure}[h]
	\includegraphics[scale=1]{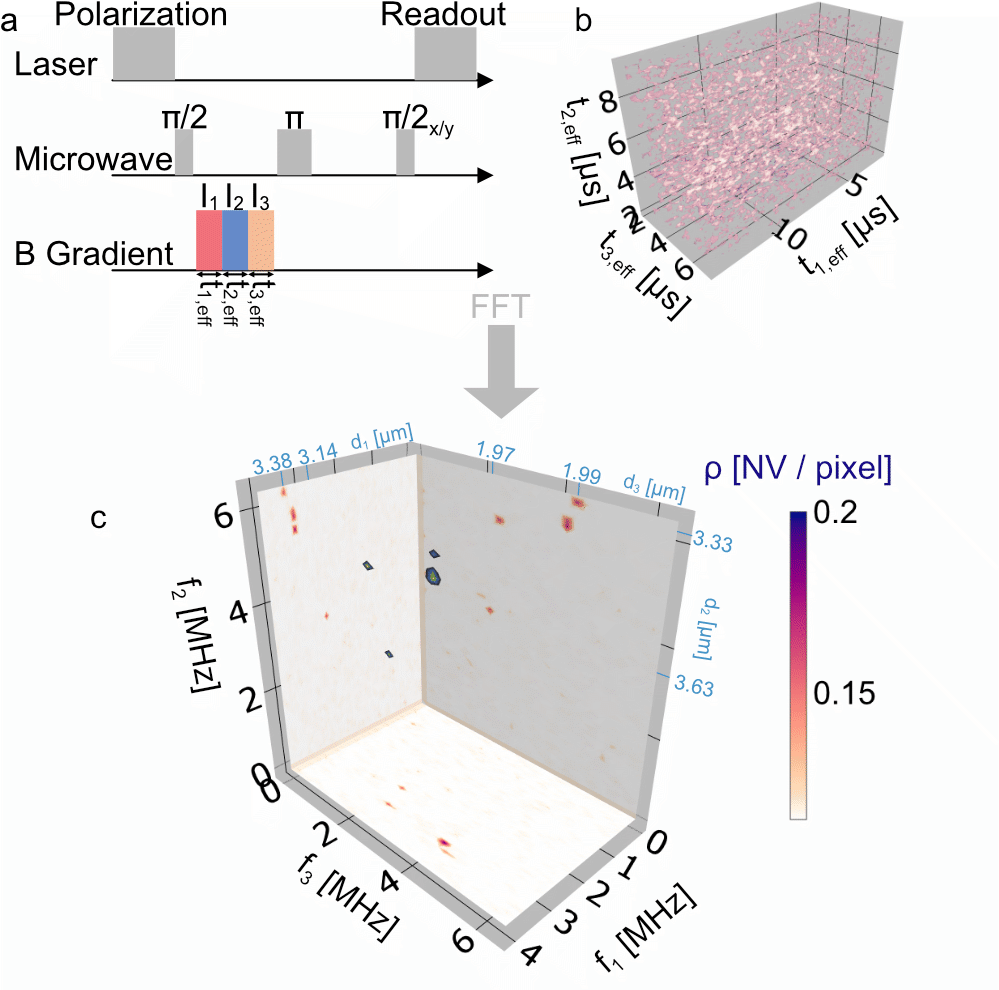}
	\caption{\label{fig:exp2} Three-dimensional magnetic resonance tomography. (a) Pulse sequence. The sequence of Fig.~\ref{fig:exp1} is extended to contain three magnetic gradient pulses from different wires. (b) Time domain data recorded from the sequence of (a) ending with $\pi/2 _{x}$. (c) Three-dimensional Fourier transform of the data in (b). The plot shows the square of the absolute value (spectral power) of the Fourier transform. $d_1, d_2, d_3$ denote the distance to the respective wire (see supplementary). The bottom, left, and back faces show projections of the 3D data.}
\end{figure}

We now extend this one-dimensional magnetic resonance tomography to three-dimensions, employing the three magnetic field gradients provided by the currents $I_1, I_2, I_3$ of our device. Note that these gradients are linearly independent if the focal spot of the microscope is placed a few micrometers below the plane of the U-structure. 


During the Hahn Echo sequence the pulses of these three currents are applied consecutively (see Fig.~\ref{fig:exp2} (a)). Since the accumulated phase will just add up linearly, the resulting spin signal is given by:
\begin{multline*}
\langle \hat S_z\rangle (t) = \Bigg{(}1 + \sum_{\vec x_{\textit{NV}}}  \cos(\omega_{I_1}(\vec x_{\textit{NV}}) t_{1} + \\ \omega_{I_2}(\vec x_{\textit{NV}}) t_{2} + \omega_{I_3}(\vec x_{\textit{NV}}) t_{3}) \Bigg{)}/2
\end{multline*}
(assuming a trailing $\pi_x/2$ pulse).


In analogy to one-dimensional tomography the set of ($\omega_{I_1}(\vec x_{\textit{NV}})$, $\omega_{I_2}(\vec x_{\textit{NV}})$, $\omega_{I_3}(\vec x_{\textit{NV}})$) can be recovered from the three-dimensional time domain data (Fig.~\ref{fig:exp2}(b)) by a 3D inverse Fourier transform, forming a three-dimensional image. 
We note that this resulting image is distorted because the gradients, while linearly independent, are not fully orthogonal. While this distortion can in principle be corrected by computing and inverting the exact spatial distribution of the frequency shift $\omega_I(\vec x)$, the raw result of the Fourier transform is still a true three-dimensional image. 
This resulting image reveals individual NV centers in the diamond. Only NV centers within the confocal volume of the microscope can be imaged. For the given NV density $5-15$ centers are expected to be seen.





\begin{figure*}[ht]
    \includegraphics[scale=1]{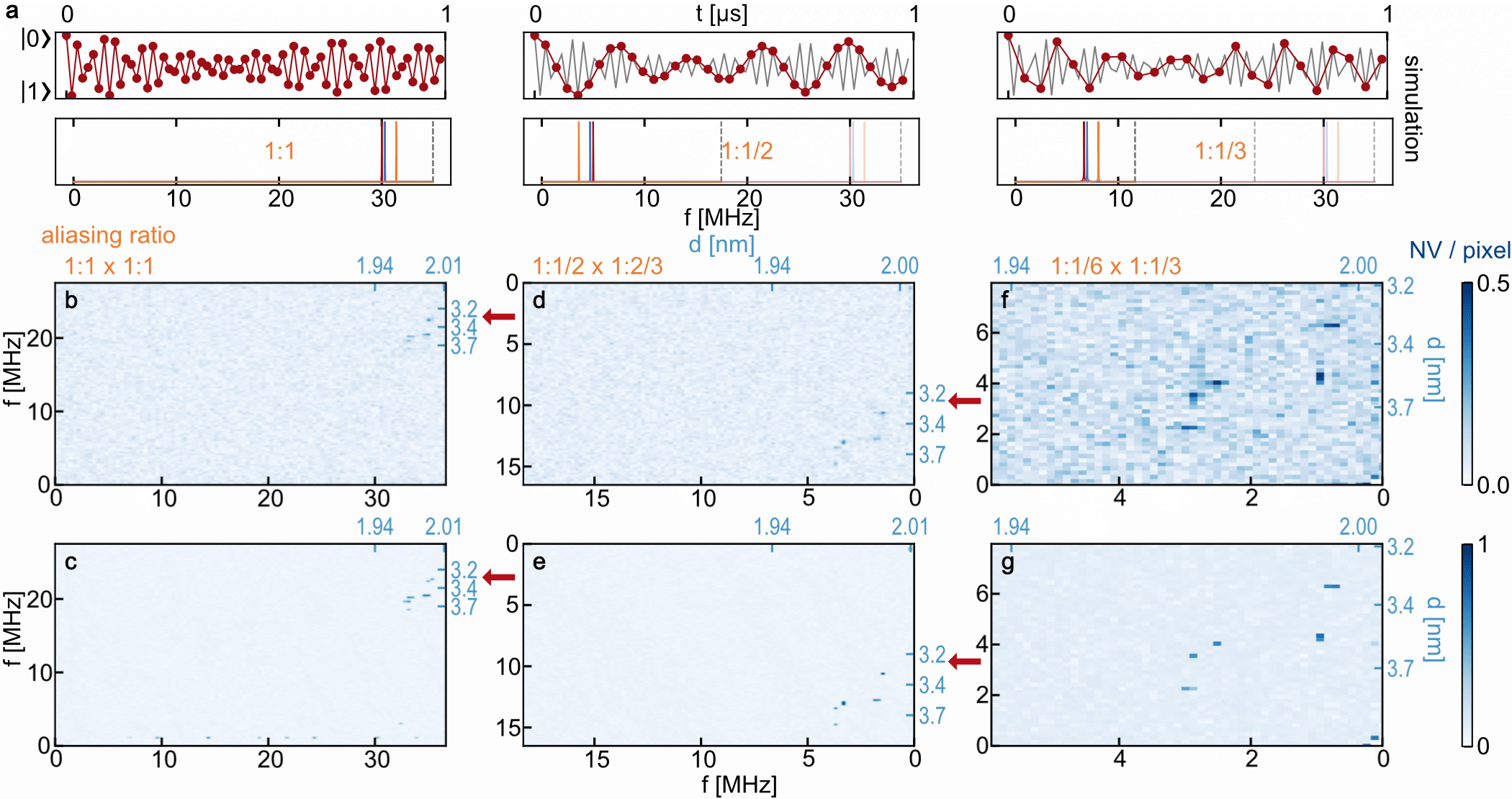}
    \caption{\label{fig:exp3} Aliasing magnification and speed-up of Fourier magnetic imaging. The plots in (a) show a simulated signal and its Fourier transform. Going through the columns from left to right the sampling rate is reduced resulting in a slower oscillation (red trace) and a lower Nyquist frequency (black-dashed line). Aliasing around the closest even multiple of the undersampled Nyquist frequency shifts the signal to a window close to $f=0$, but does not change its shape. The  Nyquist frequency of the undersampled signal has to be chosen large enough to cover the entire signal bandwidth. The signal is shifted to negative frequencies, and hence appears flipped in a frequency axis using $|f|$, if it sits on the left of the closest even multiple of the Nyquist frequency. Panels b-g display measured 2D images of NV centers acquired by taking the FFT of the time domain signal (b, d, f) or by doing an L1 minimization of the time domain signal (c, e, g). (b,c) The Nyquist frequencies for each gradient direction (x and y axes) were set to be larger than the highest frequency in that direction (no aliasing). For (d,e) and (f,g) the measurement was done using an aliased grid, the aliased factor for each direction can be read in orange above FFT panel of each measurement. In (d,e) the undersampling parameters are such that the signal is flipped in both axes. For (f,g) the signal is flipped in the horizontal axis, but remains unflipped in the vertical axis. The aliased measurement (f,g) reduces the acquisition time by a factor of 10.}
\end{figure*}

One challenge of such multidimensional measurements is the large number of required data points in Fourier space, e.g. $10^6$ points in Fig.~\ref{fig:exp2}. This number can be reduced by compressed sensing, where only a subset of points is acquired, and the image is reconstructed by numerical techniques like $L_1$ minimization, exploiting the {\em a priori} knowledge that the signal is a sparse set of discrete points \cite{arai15}. 
Interestingly, our experimental setting allows for another compressed sensing approach. It does not require elaborate numerical reconstruction and exploits a different kind of {\em a priori} knowledge: that the signal is restricted to a narrow region of interest, i.e. a narrow band in frequency space. In this special case, we can implement an effective "zoom" into this region of interest by undersampling the signal in the time domain, lowering the amount of data points. Undersampling leads to aliasing of the signal in frequency space. For suitable parameters, this will shift the signal frequency band to a contiguous low frequency window where it can still be recovered by the inverse Fourier transform, effectively implementing a zoom. 
We demonstrate a proof of concept of this idea in Fig.~\ref{fig:exp3}. The simulated one-dimensional time and frequency domain plots (left part Fig.~\ref{fig:exp3} (a)) display a limited frequency band. When the time-domain signal is undersampled, i.e. sampled at a rate that the Nyquist frequency $f_{\textit{Nyq}}$ is smaller than the highest signal frequency, any signal at $f > f_{\textit{Nyq}}$ will be aliased to a frequency 
\begin{equation}
\label{equation of folding}
    f_{\textit{obs}} = |f - 2N\cdot f_{\textit{Nyq}}|~,
\end{equation}
where $N$ is the integer minimizing $|f-2Nf_{\textit{Nyq}}|$. In (Fig.~\ref{fig:exp3} (a), middle plot), the signal band (around $f\approx 30~$MHz) is close to twice the Nyquist frequency ($2\cdot f_{\textit{Nyq}}= 35~$MHz) and hence aliased to a region close to $f=0~$MHz. Note that the aliasing involves mirroring of the signal band, because the signal is at a lower frequency than the closest even multiple of $f_{\textit{Nyq}}$. 

We show this concept acquiring three separate two-dimensional measurements (Fig.~\ref{fig:exp3} (b-g)), which display NV centers in a limited region of interest (upper right corner in Fig.~\ref{fig:exp3} (b,c)), defined by the confocal volume of the microscope . This process of undersampling and aliasing implements a zoom into the region of interest (Fig.~\ref{fig:exp3} (d-e)). 
Note that this process requires the signal to be confined to a limited window of frequencies. 
Since frequencies equal to an integer multiple of $2f_{Nyq}$ will appear at the same $f_{obs}$ (Equation.~\ref{equation of folding}), signals outside the zoom window will fold back into the signal of interest. To prevent contamination of the resulting image, the signal should be bandpass limited, i.e. values outside the frequency range of the signal should be zero and the Nyquist frequency should not fall below the bandwidth of the NV signal spectrum.
For a suitable parameter choice of the undersampling, a zoom can be achieved that exactly covers the region of interest (Fig.~\ref{fig:exp3} (f,g)), allowing for the acquisition of a full image with a greatly reduced number of data points. In the specific example (Fig.~\ref{fig:exp3} (f,g)), the two dimensions are undersampled by a factor of $6$ and $3$, reducing the number of data points by a factor of $18$, i.e. more than an order of magnitude. Note that reconstruction and visualization are still feasible by an inverse Fourier transform (Fig.~\ref{fig:exp3} (f)). $L_1$ minimization is not required for reconstruction, but can still be implemented to improve the quality of the image and/or further reduce the number of data points required (Fig.~\ref{fig:exp3} (c,e,g)). 
We finally note a constraint of the technique. The undersampled data points have to be placed equidistantly in time, since any jitter or chirp will lead to a spectral broadening in the frequency-space image. Since a variation of the gradient current over the duration of a pulse is indistinguishable from a variation in timing, this also places higher demands on the constancy of currents, i.e. the requirement of rectangular current pulses discussed above.

\begin{figure}
	\includegraphics[scale=1]{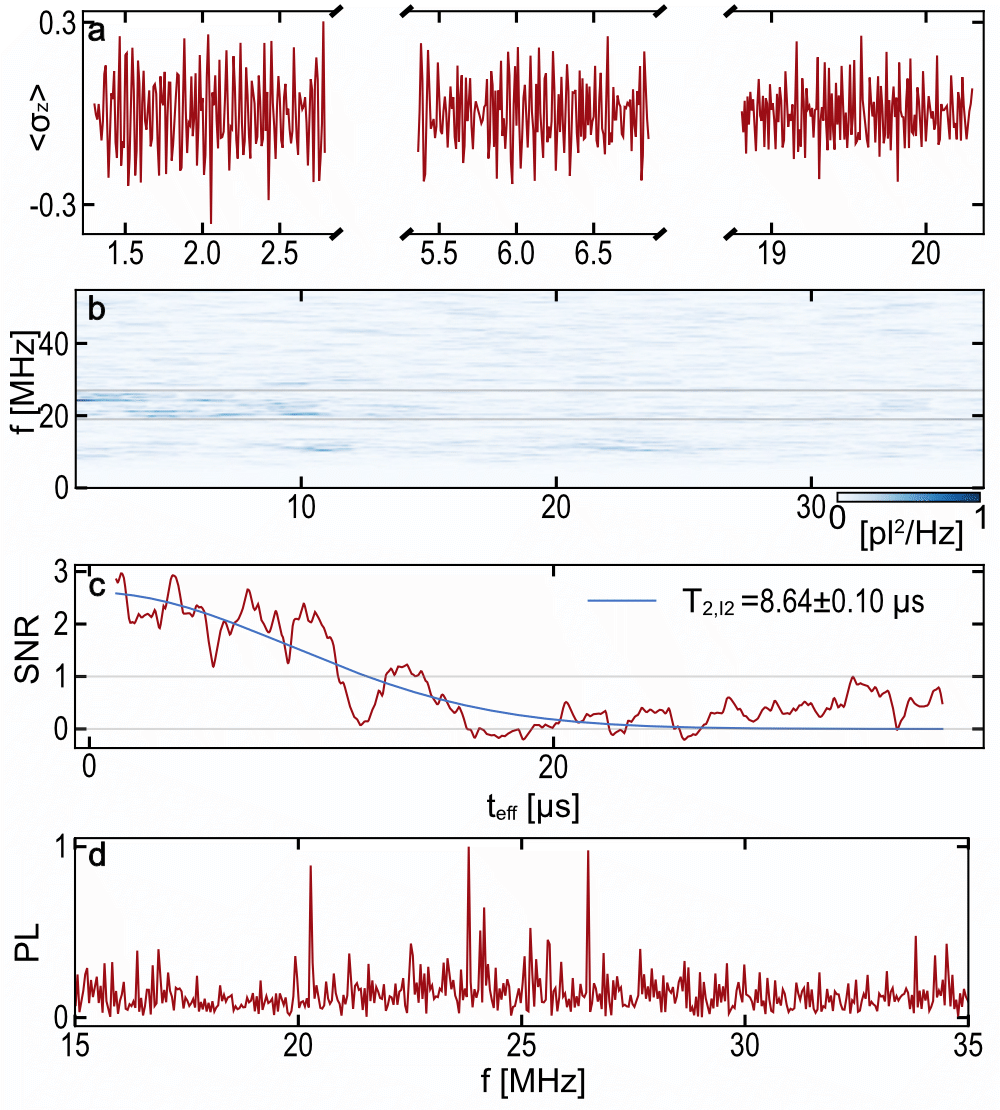}
	\caption{\label{fig:exp4} Benchmarking of the spatial resolution for I$_{2}$. (a) time-domain signal of a one-dimensional tomography (sequence of Fig.~\ref{fig:exp1}(b)). Excerpts at different time windows are shown. (b) Spectrogram (windowed Fourier transform) of (a). The signal produced by the NV centers decays over a timescale of $\approx 10~$\textmu s. (c) Signal-to-noise ratio of (b), computed by integrating the power in the signal window marked in (b) and referencing it to the noise observed outside this window (see supplementary). (c) A Gaussian fit to the data yields a decay timescale $T_{2, I_2} = 8.64\pm0.1~$\textmu s. (d) Fourier transform (absolute value) of the time domain signal in (a). 
 } 
\end{figure}

We finally analyze the spatial resolution that is achieved in our measurement. This is defined by the magnitude of the magnetic field gradient, and the spectral resolution of the spectroscopy. The frequency resolution of Fourier-transformed data is given as the inverse of the length of the time domain signal. Analogous to that, the frequency (and thus spatial) resolution of our magnetic resonance tomography depends on how long we can make the gradient pulse length and still observe an oscillatory spin signal. The longest usable pulse is limited by the fact that the spin signal decays over time on a timescale of $\approx 10~$\textmu s (see e.g. Fig.~\ref{fig:exp4} (a-b)) because of shot-to-shot fluctuations of the gradient currents, which result in the decoherence of the NV centers. Denoting the timescale of this decay by $T_{2,I}$ (i.e. the coherence time in the presence of the gradient current), the frequency resolution is 
\[ \Delta f = \frac{\sqrt{2}}{\pi T_{2,I}} \]

where $\Delta f$ denotes full width at half-maximum (FWHM) of the peak in frequency space. See supplementary information for explanation of the $\frac{\sqrt{2}}{\pi}$ factor.
We extract $T_{2,I}$ from a long 1D tomography data set, extending to several multiples of $T_{2,I}$. We calculate the Fourier transform for short time window and "slide" this window over the whole range of the time domain signal. The resulting spectrogram (Fig.~\ref{fig:exp4} (b)) shows the evolution of the NV spectrum with increasing gradient pulse lengths. The signal from a single spin appears as a horizontal line in a specific frequency band. The decaying power of the signal with increasing time in this band defines the SNR over the measurement (see supplementary). We Fit the SNR curve  (Fig.~\ref{fig:exp4} (c)) with a Gaussian ($e^{-t_{\textit{eff}}^2/2T_2^2}$) to obtain $T_{2,I}$. For the data of Fig.~\ref{fig:exp4} we thus arrive at a coherence time of $T_{2,I_2}=8.64\pm0.1~\mu$s. Combined with a gradient of $||\nabla \omega(\vec x)||/2\pi = 6.34~$kHz/nm, obtained from a numerical simulation of the gradient field (see supplementary), this corresponds to a spatial resolution of $\sigma_{x,I_2}=8.22 \pm 0.10~$nm. Similarly, we obtain $\sigma_{x,I_1}= 5.99 \pm 0.07~$nm and $\sigma_{x,I_3}= 14.47 \pm 0.50~$nm for the other two gradient currents. 
This resolution could be limited by several effects. First, shot-to-shot fluctuations of the current could shorten $T_{2,I}$. We try to suppress this by hardware integrating every single current pulse (see above) and applying post-processing corrections, but this process is equally limited by electronic noise at a lower level. Second, a spatial drift of the current path between successive experimental repetitions can equally lead to a decrease of $T_2$. A spatial drift could arise from heat expansion of the diamond and the conductors, but an expansion on the level of $10^{-3}$ would require a temperature difference of $\approx 1000~$K. Which seems unlikely. A drift of the current path within the conductor, due to local heating appears more reasonable. Intriguingly the product of $\omega_{\textit{NV}} T_2$, i.e. the relative stability of the gradient field differs between the three wires. This tentatively suggests that spatial drifts of the current in the wires are the limiting factor rather than electrical fluctuations, which would be expected to be the same in all wires.


In summary, we have demonstrated Fourier-accelerated 3D imaging of single spins with nanoscale resolution. We have also presented a compressed sensing scheme, which exploits a limited field of view, rather than sparseness of the data. Our experiments demonstrate that resolution in the sub-10$~$nm range can be achieved by switchable magnetic field gradients.

While our experiment has been performed on NV centers inside the diamonds, the device and measurement technique could equally be applied to dark spins outside of the diamond. Here a single NV center would merely serve as a detector to enable electron/nuclear spin spectroscopy on spins, while the entire process of imaging could be performed by the device presented here. Our compressed sensing technique of "Fourier zooming" will be especially advantageous in this setting where all the spins are confined to the nanoscale detection volume of a shallow NV center.
Such a direct 3D imaging technique could image an arbitrary number of spins and constrain inter-spin distances larger than $80~\textup{\AA}$, which is not possible by current electron spin resonance spectroscopy. Shrinking the structure by one order of magnitude would even push the resolution into the range of $\textup{\AA}$. Notably the $T_2$ of established spin labels is sufficiently long for the spectroscopy presented here\cite{soetbeer18}. 
\\

This work has been supported by the Deutsche Forschungsgemeinschaft (DFG, grants RE3606/1-2, RE3606/3-1 and excellence cluster MCQST EXC-2111-390814868, SFB 1477 “Light–Matter Interactions at Interfaces” (Project No. 441234705)) and the European Union (ASTERIQS, Grant Agreement No. 820394). Y.H. acknowledges financial support from the China Scholarship Council. The authors acknowledge the help of Regina Lange and Anja Clasen with taking the SEM picture and helpful technical discussions with John Marohn.

\bibliography{bibliography.bib}

\end{document}


\title{Supporting Information}
	\maketitle

\section{gradient simulation}

\begin{figure}[h]
\includegraphics[scale=1]{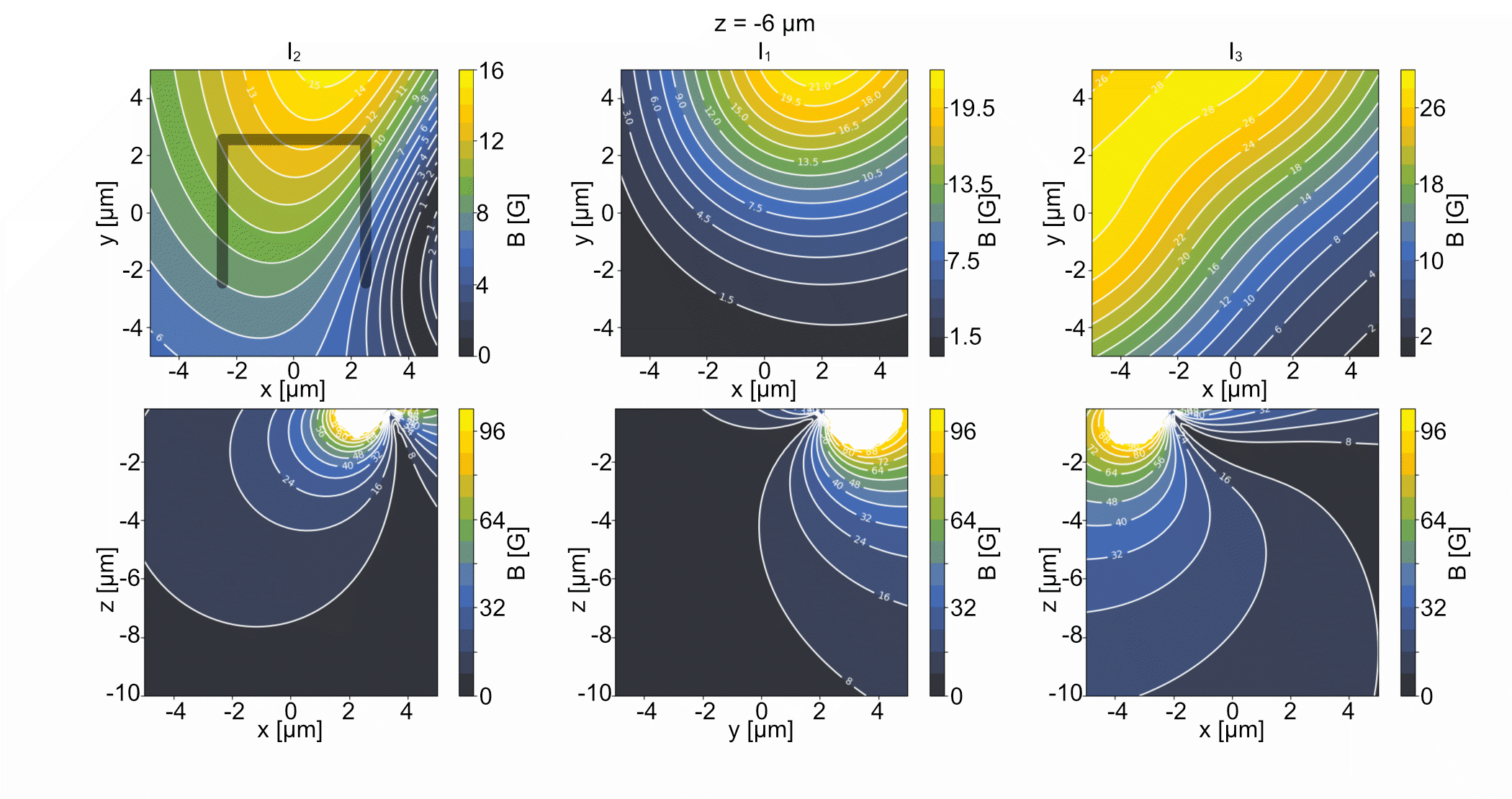}
\caption{\label{fig:sup1} Simulated field strength in units of Gauss. The first row of xy-plots from left to right shows the magnetic field generated by $I_2, I_1, I_3$, respectively, at a depth of $6~$\textmu m below the plane of the wires. The plot for $I_2$ on the left shows the location of the U microstructure as a grey overlay. The second-row present projections to the xz, yz, and xz-plane of the field from currents $I_2, I_1, I_3$, respectively, taken at $x=0~$\textmu m (or $y=0~$\textmu m).}
\end{figure}

The magnetic field generated from $80~$mA through the $5~$\textmu m arms of the U microstructure was simulated using Python's library "Magpylib". The simulated area was $2~$mm$\times 2~$mm large to include the influence of the larger feed lines connected to the U structure. The temperature of the wires was assumed to be constant, which is a valid assumption, because of the high heat conductivity of the diamond substrate. The $0.5~$\textmu m wide wires were divided into 10 subdivisions each to take spatial inhomogeneities into account. More subdivisions did not lead to a change of the outcome of the simulation. We then matched the simulated magnetic field strengths to the frequencies of the measured spin signal to locate the NV centers in three dimensions. This was done with a minimization algorithm, which converged to a point $6~$\textmu m below the Diamond surface. This matches the depth measured optically with the confocal microscope. Fig.~\ref{fig:sup1} displays the field value in units of Gauss at a depth of $6~$\textmu m below the wires. By extracting the magnetic field gradient and calculating the Jacobian at the point of interest, we can extract the sensitivity ($|\nabla \omega(\vec x)|/2\pi$) along the shortest distance to each of wires carrying $I_1, I_2, I_3$ to be $7.63~$kHz/nm, $6.34~$kHz/nm, $6.86~$kHz/nm, respectively.

\section{Signal-to-noise ratio (SNR)} 
The signal-to-noise ratio (SNR) is defined as the ratio of signal power to the power of the background. The power of the signal is calculated by integrating the spectrogram-band between the two gray lines seen in Fig.~4 (in the main text) and subtracting the background from it. Here, the background is calculated by integrating spectrogram-band where there is no signal and rescaled it to account for the difference in band size between the two integrated windows. The power of the noise is calculated the same as the background, where the integral of the band outside that of the signal is taken. 


\section{Resolution}

Frequency resolution of the method is limited by the decay of the signal, and, in general, is inversely proportional to the lifetime $\Delta f = k/T_2$. The constant of proportionality $k$ is given by the exact shape of decay, which in this case is assumed to be Gaussian:
$$S_z(\tau) = \cos({\omega \tau}) e^{-\frac{1}{2}{ (\frac{\tau}{\sigma}})^2}$$
Since only the $\tau>0$ half of the Gaussian envelope is observed, analytical solution for the $k$ is difficult, thus numerical simulation was performed instead. The idea of simulation is to extract FWHM of the peak in frequency space and relate it to the $\frac{1}{T_2}$, manually finding a constant until both will coincide. The code listing is given below. It was found that for the given signal model the $k=\frac{\sqrt{2}}{\pi}$
\begin{lstlisting}[language=Python, caption=Peak width simulation]
import numpy as np
import matplotlib.pyplot as plt
import scipy.signal 

t = np.linspace(0, 2000, 50000) # us
f_max = 1 / (t[1] - t[0]) # MHz
T2_s = np.arange(2, 20) # T2 time
freqs = np.linspace(0, f_max, len(t)) # fft frequencies
df = freqs[1] - freqs[0]

def model(sigma): # signal model, cosine weightes by gaussian envelope
    return np.cos(10*np.pi*t) * np.exp((-1/2)*((t/sigma)**2))

widths = []
for t2 in T2_s:
    y = model(t2) # we define T2 as std of gaussian envelope
    spectrum = np.abs(np.fft.fft(y))[:len(t)//2]**2
    width = scipy.signal.find_peaks(spectrum, width=(None, None))[1]['widths'][0]
    widths.append(width)
    
widths = np.array(widths) * df # need to multiple by df to get MHz from pixels
k = np.sqrt(2) / np.pi # educated guess

plt.plot(T2_s, k/T2_s, "o-")
plt.plot(T2_s, widths, "o-")
plt.xlabel("T2, us")
plt.ylabel("Peak width, MHz")
\end{lstlisting}